\documentclass[nobalancelastpage,aps,showkeys,showpacs,nofootinbib,
  twocolumn,a4paper,10pt
]{revtex4-1}

\usepackage{amsmath}
\usepackage{amssymb}
\usepackage{url}
\usepackage[dvipdfmx]{graphicx}
\newcommand{\hh}{%
  $\mathrm{H}_2$} 
\newcommand{\tp}{_\mathrm{tp}} 
\newcommand{\kb}{k_\mathrm{B}} 
\newcommand{\tc}{T_\mathrm{c}}
\newcommand{\ec}{E_\mathrm{c}}
\newcommand{\myrule}{%
  rule \eqref{sign}
}
\newcommand{\Myrule}{%
Rule \eqref{sign}
}

\begin{document}
\title{%
  For Noble Gases, Energy is Positive for the Gas Phase, Negative for the Liquid Phase
}

\author{ASANUMA Nobu-Hiko}
\email{phys.anh@z2.skr.jp}
\homepage{http://z2.skr.jp/phys/}
\affiliation{No research affiliation}

\begin{abstract}
  We found from experimental data that for noble gases and \hh, the energy $E$ is positive for the gas phase, and negative for the liquid, possibly except the small vicinity of the critical point, about $(1- T/\tc) \lesssim 0.005$. The line $E=E_c$ in the supercritical region is found to lie close to the Widom line, where $E_c$ is the critical energy.
\end{abstract}

\pacs{64.70.fm, 
  64.60.Ej, 
 45.50.Jf 
}

\keywords{%
  energy, vapor, liquid, bound state, many-body system, critical point
}

\maketitle

\section{Introduction}

What distinguishes the gas phase from the liquid phase? For noble gases, we declare that it's the sign of the energy $E$:
\begin{equation}
  \label{sign}
  E > 0 \,\, \text{for gases,} \, \, E < 0 \, \, \text{for liquids}.
\end{equation}
It's because a gas is an unbound state, so it cannot be $E<0$; otherwise, it would condense. Similarly, if $E>0$, a liquid could not be a bound state. Of course, this argument is mean-field theoretic---at best. Actually it's too crude, and we cannot justify it logically.

It is however correct, as we will see from experimental data. To be precise, it may be violated in the very small vicinity of the critical point; according to the used data, it does not hold only for $t < 0.006 $ along the saturation curve (gas-liquid coexistence curve) for 4 noble gases, where $t:= 1 - T/\tc$ is the reduced temperature, $T$ the temperature, and $\tc$  the critical temperature.
For the extreme case of \hh,\footnote{
  For $T < \tc$, the internal degrees of freedom of \hh~can be ignored, as is explained in section \ref{section-hh}.
}
the violation is only for $t < 0.001$. See figure \ref{eSign} below.

We report that \myrule is corroborated experimentally, but notice that it is more of theoretical value. For example, it can raise interest for dynamical system theory. We cannot determine from the experimental data if \myrule is exact for real fluids, but if it really is, the critical energy $\ec$ is $0$, or equivalently $K = |U|$ at the critical point, where $K$ is the kinetic energy and $U$ the potential energy. This strongly indicates a symmetry. Even if it is not exact, we conjecture that there is a symmetry, and it is weakly broken. More will be discussed in section \ref{section-discussion}.

\Myrule somehow seems to have been unnoticed despite its simplicity and decisive power.\footnote{
  There is a sole exception, ref \onlinecite{elsner}, but the arguments of this paper do not make sense. In section 2, they ``prove'' that the zero of the internal energy in thermodynamics is not arbitrary. This is of course absurd. What's really proven is that it's not possible to assign arbitrary zeros to each of subsystems of one entire system. In spite of this assertion, they do not define the zero of the energy, nor does it mention molecule's internal excitations. Then in section 4, they assume that the sign of the energy cannot change within one phase, and concludes that $E > 0$ for a gas, $E =0$ at the critical point, etc.}
In the article in 2012 titled ``What separates a liquid from a gas?'', \cite{physToday-frenkelLine} it is not mentioned. It is not found in recent textbooks of statistical physics \cite{pathria-beale:2011:3ed,kardar-stat-phys,sethna,chaikin-lubensky,goldenfeld} nor in liquid theory textbooks. \cite{schirmacher-liquid,hansen-mcdonald-simple-3ed}

Looser explanations like ``$K \gg |U|$ for gases, $K \ll |U|$ for solids, and $K \approx |U|$ for liquids'' are on the other hand common. Ref. \onlinecite{stishov} studied these relations a bit more further for van der Waals fluid,  and heuristically obtained the estimate that $|U| / (\kb T) \approx 0.9$ at the critical point, and concluded that far from the critical point the relations $\kb T \gtrless |U|$ are  good estimates for the liquid and gas phases, but not to the precision we give.

This letter is organized as follows. In section \ref{section-method}, the  experimental data we used and the theory are explained. In section \ref{section-result}, the result is stated. In section \ref{section-wl}, we try to extend \myrule to the supercritical region, and we draw a qualitative conclusion that the line $E=\ec$ lies close to the ``Widom line'',\cite{widomLine} the line of the maxima of $C_p$, the constant-pressure heat capacity. Section \ref{section-discussion} gives discussion and outlook. Section \ref{section-conclusion} is the conclusion.

\section{Method}
\label{section-method}
\subsection{Cited experimental data}
As ``experimental'' data, we rely on NIST Chemistry WebBook data on fluids (hereafter ``WebBook''). \cite{nist-formatted} In fact they are not true experimental data, but the output of the program ``REFPROP'' which computes model equations. Their parameters are fit to the results of experiments done in various conditions, ranging from low to high temperature and pressure, and near and far from the critical point. In addition, models differ from substance to substance. Thus an accurate error estimate is not available. It is only stated that ``These equations are the most accurate equations available worldwide.'' \cite{refprop-faq-formatted}

The lower bound of the temperature at which they provide data is $T\tp,$ the triple point temperature, and for He, the $\lambda$-point temperature. They provide data along the saturation curve, in addition to isotherm, isobar, etc.

They provide data on 75 fluids. All noble gases except Rn are included.

\subsection{Definition of $E=0$}
To define the zero of the energy, for noble gases we safely ignore internal states, i.e. thermal excitation of electrons. (The first excitation energy of, for example, He is about 20eV, and that of Xe is 8.3eV.%
)
In dilute limit, all fluids become an ideal gas. Therefore we naturally define that $E = 3/2 N\kb T$ in dilute limit. Here, $N$ is the number of atoms.

WebBook provides the data on various thermodynamic properties, and we in particular need those on the internal energy. The zero of the internal energy is arbitrary, and in WebBook it depends on the kind of fluid.\footnote{
  For most fluids it says that the origin of $E$ is taken at $T=273.15$K for ``saturated liquid'', but for not few fluids it is in the supercritical region, and this explanation is dubious.
}
To interpret the energy of WebBook, we determine the zero of $E$ from the value at $T=\tc, p = 0$Mpa for each fluid, where $p$ is the pressure. (WebBook indeed provides data down to 0MPa,  probably extrapolated.) This choice of $T$ is arbitrary, and does not matter. At these points, $ | C_v / (3/2N\kb) - 1| < 10^{-3}$ for all available substances, where $C_v$ is the constant-volume heat capacity. So they can be reliably thought as  dilute limit.

\subsection{Inclusion of \hh}
\label{section-hh}
We also examine the behavior of \hh \, because for $T \le \tc$ internal excitations are almost ``frozen'' and can be ignored. (According to WebBook, $|C_v/(3/2N\kb) - 1| = 3 \times 10^{-5}$ at $T=\tc, P=0$MPa.) It's because hydrogen is an exceptional molecule by having the large moments of inertia. (This is not true even for D$_2$, deuterium, for which $C_v(\tc) / (3/2N\kb) = 1.12$ at 0MPa.)

From WebBook it's not clear if it is true equilibrium hydrogen, or ``normal hydrogen'', i.e. the 3:1 mixture of orthohydrogen and parahydrogen.
If it is normal hydrogen, an orthohydrogen molecule should be considered as stable, not an excited state of parahydrogen. So still the zero of the energy is determined as $E=3/2N\kb T$ at $p = 0$MPa, where $N$ is the number of molecules.

\subsection{Other Comments}
Helium is to some extent a quantum fluid on the saturation curve, since $\lambda^3 \rho = 0.57 \sim 1$ at the critical point, where $\lambda$ is the thermal de Broglie wavelength and $\rho$ is the number density.
But it's common to both classical and quantum mechanics that boundness is determined by the sign of the energy, so it is not necessary to modify \myrule for this case.\footnote{
  In quantum mechanics, bound states with positive energy is possible for systems. See for example ref. \onlinecite{Ballentine}, sec. 10.4. But it is only for potentials which satisfy special conditions, and we ignore such cases.
}

\Myrule should apply not only to pure substances but also to mixtures, as long as there is the natural definition of the origin of the energy, namely $E = 3/2N\kb T$  in dilute limit. Normal hydrogen falls into this category.

\section{Result}
\label{section-result}
\begin{figure}
  \includegraphics[width=8.5cm]{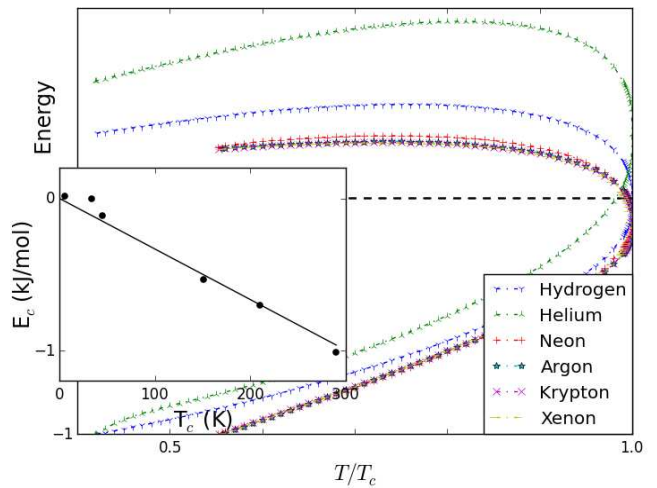}
  \caption{(Color online) The energy of 5 noble gases and H$_2$ on the saturation curve. The upper curve is of the gas phase, and the lower of the liquid. Horizontal dashed line is for $E=0.$ The curves are so scaled that $E = -1$ for the liquid phase at the triple point. Inset shows the $\ec$ and $\tc$ for these 6 fluids, and the line is for $E = -0.4N\kb T$.}
   \label{eSign}
\end{figure}

In figure \ref{eSign} we show the WebBook data of the energy of gas and liquid on the saturation curve, for 5 noble gases and \hh. The energy in the plot is so scaled that the energy of liquid at the triple point is $-1.$

As it can be seen, \myrule is satisfied  except the neighborhood of the critical point. For He, the violation of \myrule happens for $t < 0.02,$  for Ar, Kr, and for Xe $t < 0.006,$ and for Ne $t<0.005$. In the extreme case of \hh, it is only for $t < 0.001.$

Considering the inherent uncertainty of WebBook data and experimental difficulty, this agreement is remarkable and cannot be accidental. We conclude that \myrule is: ``Correct, possibly except very narrow regions near the critical point.'' However, we cannot determine quantitatively the region where \myrule does not hold, because of the lack of the error estimate in WebBook.

Possibly except an area close to the critical point, we are sure that \myrule is correct not only on the saturation curve, but in a very wide range of $p$ when $T\tp < T < \tc$. It's because heat capacity is positive, and on isotherm $\partial E / \partial p < 0$ if the pressure is not too high (but if the pressure is that high probably the system crystallizes). We are not sure for the gases in the region  $T<T\tp$ for which WebBook doesn't provide data.

We also note that $E$ is always $ > \ec$ for the gas phase, and $< \ec$ for the liquid phase on the saturation curve according to WebBook.

\subsection{Critical energy}
To assess $\ec$, we also  plot $\ec$ and $\tc$ of the same 6 fluids in the inset of figure \ref{eSign}. (Remember an error bar is not available.) We also draw the line $E=-0.4N\kb T,$ which is simply ``fit by eye.'' The agreement of this line with the experimental data looks good, so we're tempted to say that $\ec$ is indeed $\approx -0.4N\kb\tc \ne 0$, but we avoid to draw any conclusion.

\section{The line $E=\ec$ in the supercritical region and the Widom line}
\label{section-wl}
\subsection{Introduction}
\label{section-widom-intro}
We can not tell if $\ec$ is exactly $=0$, but the question if the line $E=\ec$ is still meaningful in the supercritical region is natural, possibly representing a crossover, dividing liquid-like and gas-like behavior. In fact lines of such crossover are already proposed, dubbed the ``Frenkel line''\cite{frenkel} and the ``Widom line''.\cite{widomLine} Actually we feel that the arguments on the Frenkel line are more convincing than those on the Widom line, but we here compare the $E=\ec$ line with the Widom line because of the data availability.

The Widom line is defined as the line of sharp maximum of $C_p,$ the constant pressure heat capacity, in the supercritical region, starting from the critical point. More precisely, the $C_p$ divergence at the critical point does not form a round peak, but on each isothermal and isobaric line near the critical point, a sharp $C_p$ maximum exists. By connecting those maxima, a ``ridge'' is formed, and it is the Widom line. It is also characterized as the collection of maxima of various thermodynamic response functions. Even though the validity of the Widom line notion is questioned,\cite{wlDispute} there is no problem as long as we consider an area close enough to the critical point.

\subsection{Result}
\begin{figure}
  \includegraphics[width=8.5cm]{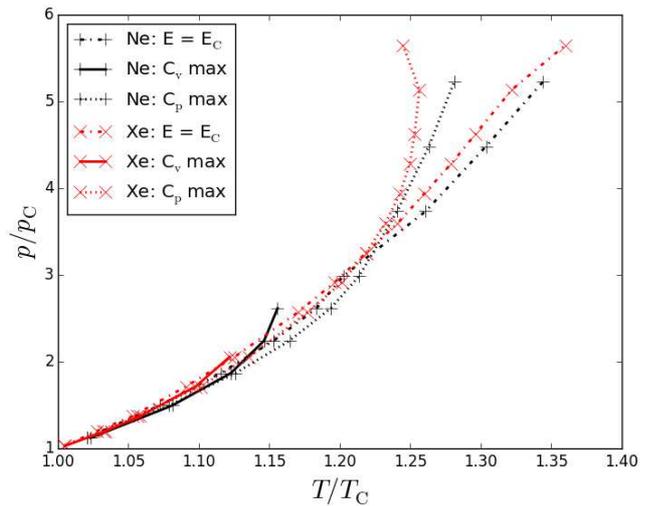}
  \caption{(Color online) The lines of $C_v$ and $C_p$ maximum and $E=\ec$ in the supercritical region for Ne and Xe. For high enough temperature, the $C_v$ maximum lines disappear so they're not plotted.}
  \label{wlFig}
\end{figure}

We plot in figure \ref{wlFig} the lines $E=\ec$, $C_p$ maximum, and also the maximum of $C_v$, the constant volume heat capacity, for Ne and Xe.

Our qualitative conclusion is that the line $E=\ec$ runs near the Widom line, in the region of low enough temperature where the Widom line can be recognized without ambiguity.

When the system moves far away from the critical point, $C_v$ maximum disappears, and the Widom line may not be well-defined there. In that region, the line $E=\ec$ departs from the $C_p$ maximum line.

We gave the plots of Ne and Xe, but our result applies to \hh, Ar and Kr, too. We cannot assert anything on He: Data close enough to the critical point are not provided by WebBook; for the region with data, the line $E=\ec$ and $C_p$ maximum do not agree well, and $C_v$ maximum cannot be observed.


\section{Discussion and outlook}
\label{section-discussion}
Rule \eqref{sign} which we judge almost correct, raises many questions. First of all, is it exact? Computer simulations should prove it; rather, a disproof will be easier than a proof---experimental verification will be difficult, because of finite-size effect and the presence of gravity.\cite{finite-size-criticality,gravity-rmp}  If correct, it must be so for any interactions which have the critical point and the natural definition of $E=0$, independent of dimensionality. (Even though the physics of noble gases  is usually thought to be well described by Lennard-Jones potential, the contribution of the three-body forces has to be taken into account to reproduce the third virial coefficient of real noble gases.\cite{cite-cusack})

If \myrule is not exact, why is its breakdown limited to the very small region near the critical point?  The equation $\ec=0$ can still be used as the mean-field, zeroth-order value, but how can corrections be calculated?  Are there any system for which exactly $\ec =0$?

If \myrule is exact, $\partial E / \partial N \to 0$ in  thermodynamic limit at the critical point.
Intuitively $E$ being $=0$ is the edge of boundness, and is also the point where a dimensionful constant vanishes, so it seems to be related to the scale invariance of the critical point. However the condition of $E=0$ is not sufficient, since the line of states $E=0$ does exist in the supercritical region too. \Myrule also means $K = |U|$ at the critical point. This strongly indicates a symmetry, directly connecting $K$ and $U$, aside from the scale invariance. Even if not exact, we can say there must be an approximate symmetry. What symmetry is it precisely? How is it related to the scale invariance?

As we cautioned, \myrule is very rough. For example, it completely ignores the formation of atomic clusters. It also treats the energy from the viewpoint of mechanics, but the energy of a fluid is a thermodynamic quantity,  the (canonical) ensemble average, which is not conserved. Definition of boundness is very involved, if ever possible, for many-body systems. At the same time, treatment in microcanonical, or dynamical system theory may be possible.

Yet, its incisive simpleness allows a clear understanding, or new definitions of gas and liquid. For example, consider the solution of  solute A and  solvent B without internal degrees of freedom. Then it can be said that A is gaseous inside the solution, and B is liquid. Let us write the Hamiltonian $H$ as:
\begin{equation}
  H = K_A + K_B + U_{AA} + U_{BB} + U_{AB},
\end{equation}
where $K_A$ is the kinematic energy of A particles, $U_{AB}$ is the potential between A and B particles, and so on. Now integrate out B's variables. Then we obtain the effective Hamiltonian $H_{\mathrm{eff}}$ which looks like:
\begin{equation}
H_{\mathrm{eff}} = K_{A\mathrm{eff}} + \sum_i U_i,
\end{equation}
where $U_i$ is the $i$-body effective potential, which is $:=0$ in dilute limit. What \myrule tells is that $\langle K_{A\mathrm{eff}} \rangle+ \sum_{i \ge 2} \langle U_i \rangle > 0,$ where the bracket is the thermal average, because A is gaseous. In addition since the whole system is liquid, $\sum_{i \ge 0} \langle U_i \rangle < -\langle K_{A\mathrm{eff}} \rangle< 0.$ So, $-\langle U_0 \rangle$ is similar to to the work function of metals, although in the current case the temperature is finite. If A and B demixes so that the A-rich phase and the B-rich phase coexist, then A is gaseous in B-rich phase and liquid in A-rich one, and so on. It seems almost obvious, in reality a mere heuristic though, that the critical point of binary fluid consolution belongs to the same universality class as gas-liquid's one.

A still easier example is the theory of dilute solutions, found in every textbook of thermodynamics. When the author was a student, he felt the appearance of the gas constant $R$ was sudden and absurd. ``Interaction is strong, the ideal gas has nothing to do here, no?'' It is the consequence of thermal average,\footnote{
  Thermodynamics of dilute solution can be derived within pure thermodynamics, without the need of statistical mechanics. See for example ref. \onlinecite{fermi-thermodynamics}, chapter 7.
}
but we have an alternative view. In fact, solute is a gas trapped in the solvent,
and in dilute limit, it becomes an ideal gas, because the interaction between solute molecules can be neglected. The mean free path of the bare solute molecule does not matter.

\Myrule imposes a limit on the spinodal curve, too. The spinodal curve is difficult to define theoretically. In textbooks, it is often explained mean-field theoretically as ``the'' inflection point of (the metastable branch of) the free energy. (See for example ref. \onlinecite{chaikin-lubensky}, section 8.7.3, or ref. \onlinecite{strobl}, section 3.4.3.) More careful definition is as the occurrence of negative compressibility for \emph{all} wavelengths,\cite{hrt-gl} but it still suffers from the fact that it may not be well defined due to metastability.
We know however that the supercooling of gas and superheating of liquid cannot exceed the line $E=0.$ It is only a necessary condition, but the energy of the system is always defined. At the very least it explains the existence of spinodal curves in gas-liquid transition.

We do not know how to extend \myrule for molecular fluids. Molecules have  internal degrees of freedom, namely ro-vibrational modes. Intermolecular interactions depend on the internal states, or in other words, they mix and it is not possible to define the quantity $U$ separately from internal states. In \myrule translational degrees of freedom are concerned, so to promote it to molecular fluids, we have to extract and separate them from internal degrees of freedom. It must be possible, since critical points exist also for molecular fluids, but we are clueless how to do it.

\Myrule also hints at something on the notion of cluster and percolation in lattice and off-lattice systems, which is easiest to describe from the standpoint of Monte Carlo simulation. (For an introduction see for example ref. \onlinecite{landau-binder}, section 5.1) ``Cluster algorithms'' in general update all variables in a group, called cluster, but we call it ``updating cluster'' (UC). There is another cluster, which percolates at the critical point, which we call PC. PC is used to locate the critical point in ``invaded cluster algorithm''\cite{InvadedCluster}. In Ising model, PC is the set of parallel spins which are connected. It is also generalized for example to Widom-Rowlinson model,\cite{Widom-Rowlinson,InvadedCluster:WR} but not for general fluids. UC is a subset of PC, and it has to satisfy detailed-balance. It is usually chosen to make the algorithm most efficient, but it is not necessary. Because percolation is deeply connected to criticality, the current situation where PC is lacking for general systems is unsatisfactory. Our questions are, how to define PC for general systems, and does UC have a physical meaning beyond a mere computational utility? Is it possible to define an analogue of the kinetic energy for lattice systems? By answering them, it may be possible to obtain more insight on the opaque relations between the lattice-gas models and fluids.

In physics, models, even toy models, have served to make various advances, and we inevitably pose this question: Is there any one-particle, central force system, classical or quantum, which has a phase transition at $T=\tc,$ and for $T \gtrless \tc, E \gtrless 0$? For classical cases, natural order parameters are $\langle 1/r \rangle$ and $\langle U \rangle.$

\begin{figure}[t]
   \includegraphics[width=8.5cm]{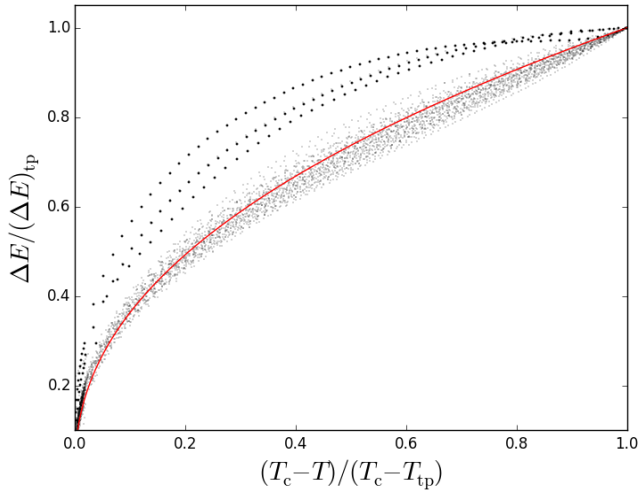}
   \caption{(Color online) The energy difference of evaporation $\Delta E(t)$ of 75 fluids \cite{nist-formatted} along the saturation curve, normalized to $1$ at the triple point $T = T\tp$. The solid line means  $t^{0.44}$ with the same normalization. The data of He, H$_2$, and D$_2$ are represented by specially thick dots, which substantially differ from others.
   }
   \label{fig-eDiff}
\end{figure}

\Myrule also suggests that the energy may be an order parameter. What we have discovered recently \cite{my-univ} is that the energy difference $\Delta E(t)$ of evaporation along the saturation curve is universal, by being well approximated by the power law $\propto t^a$, where $a \approx 0.44$, \emph{including molecular fluids}. Figure \ref{fig-eDiff} shows $\Delta E$ for 75 fluids of which data is provided by WebBook, This is surprising and uncanny, because ro-vibrational modes are diverse among substances.\footnote{
  In the figure, $\Delta E$ is normalized to 1 at the triple point. This normalization was inspired by the work by Torquato et al. \cite{torquato} which reported  similar universality for the latent heat $T\Delta S$. They  represented $T\Delta S$ by 6-term polynomials with non-universal, fluid dependent coefficients.
}
We also found that  $T\Delta S \propto t^{0.38}$. They are not critical phenomena; these two power laws apply to the entire saturation curve \emph{except an area near the critical point}, but instead \emph{down to the triple point}. The critical exponents of them and of $p\Delta V = -\Delta F$ are $\beta$.\footnote{
  $\Delta V(t)$ and $\Delta F(t)$ are differences of volume and Helmholtz energy along the saturation curve. The critical exponents of $\Delta V$ and hence of $p\Delta V$ are strictly $=\beta$. If asymptotically $p_\text{c} -p \sim t$, which is almost certain, by the Clausius-Clapeyron relation $\Delta S, T\Delta S \sim t^\beta$. Then by noting $\Delta G = 0$, the difference of Gibbs energy along the saturation curve, it is likely that $\Delta E \sim t^\beta.$
}
Here $\Delta E$ and $T\Delta S$ are per particle; contrary to our motivation, we couldn't find anything conclusive on the \emph{spatial} energy density.

We pointed out that $E$ is a quantity that can be defined purely in mechanics, without thermodynamics. But not only $\Delta E,$ but also $\Delta 1/V,$ the density difference, is an order parameter along the saturation curve, as known very well, and $V$ is a pure mechanical quantity, too. Some mysterious truth seems to be still hidden.

\section{Conclusion}
\label{section-conclusion}
We found from experimental data that for noble gases and \hh, the energy is positive for the gas phase, and negative for the liquid phase. According to the used data, this rule dose not hold for $1 - T/\tc < 0.02$ for He, for other 4 noble gases, $1 - T/\tc < 0.006$, and for \hh, $1 - T/\tc < 0.001$, along the saturation curve. The used data does not provide any error estimate. In the supercritical region, we draw the qualitative conclusion that the line $E=E_c$ runs close to the Widom line in the region not so much far from the critical point.

\bibliography{bib/phys.bib,bib/extra.bib}
\bibliographystyle{apsrev}

\end{document}